
\magnification=\magstep1

\def\\{\hfil\break}

\tolerance=2000

\def\newpage{\vfill\eject}

\def\\{\hfil\break}


\hsize 15.2truecm
\hoffset=1truecm
\parskip=4pt

\def\newpage{\vfill\eject}

\def\\{\hfil\break}

\vskip 3 cm

\noindent
\line{\hfil                                     BI-TP 93/18}
\vskip 3.0truecm

\centerline{\bf WEAK COUPLING PHASE STRUCTURE}

\centerline{\bf OF THE ABELIAN HIGGS MODEL}

\centerline{\bf AT FINITE TEMPERATURE}

\bigskip\bigskip\bigskip
 \centerline{A. Jakov\'ac$^1$ and A. Patk\'os$^{1,2}$}
\medskip

\centerline {$^1$\sl  E\"otv\"os University, Dept. of Atomic Physics}
\centerline {\sl H-1088 Puskin u. 5-7, Budapest, Hungary}
\medskip
\centerline {$^2$\sl Fakult\"at f\"ur Physik, Universit\"at Bielefeld}
\centerline{\sl Postfach 10 01 31, W-4800 Bielefeld 1, Germany}

\vskip 5.0truecm

\centerline{\bf Abstract}\medskip
{\narrower\noindent
Using the 1-loop reduced 3D action of the Abelian Higgs-model we discuss
the order of its finite temperature phase transition.
  A two-variable
saddle point approximation is proposed for the evaluation of the
effective potential. The strength of the first order case
scales like $\sim e^{3-6}$. Analytic
asymptotic weak coupling and numerical small coupling solutions are
compared with special emphasis on the cancellation of divergences.}
\newpage
\noindent{\bf 1. Introduction}
\bigskip
The observed asymmetry between matter and anti-matter in the Universe is
very probably the consequence of some out-of-equilibrium situation
in the course of its History, rather than part of the initial data.
[An up-to-date progress report on the subject is provided by Ref. [1].]
Any starting net baryon number would have been washed out by the
high-temperature B-violating processes, which are present in thermal
equilibrium even in the Standard Model of electroweak interactions.

Out-of-equilibrium state of matter
occurs in first order phase transitions. This circumstance has been
successfully associated by Kuzmin {\it et al.}[2] with known C and CP
violation and intense anomalous B-violation at high temperature in
the Standard Model to argue that Sakharov's conditions [3] for
baryogenesis can be met within a maximally known physical framework,
too. Quantitative agreement with the observed asymmetry might enforce
the extension of the particle content of the model enhancing the strength
of CP-violating effects [4].

The present attention focused on the nature and the quantitative
features of the electroweak phase transition should be appreciated
on this background. The principal tool of the investigation is the
effective potential at finite temperature.

In the usual loop-expansion propagators of all particles are evaluated
with masses extracted from the tree-level shifted Lagrangian. For small
values of the scalar field some of the mass-squares become negative,
what prohibits any clean interpretation of effective potential, the
interface features, etc. Summation of the class of "daisy-type",
infrared sensitive contributions leads to the cancellation of the dangerous
pieces [5]. A short-cut in this gradual procedure is to use the temperature
corrected effective masses from the start [6]. A complete self-consistent
treatment of the polarisation functions has been described by
Buchm\"uller {\it et al.} [7] (see also [8]), where the masses are
roots of a set of gap-equations.

The significance of the effective mass is obvious because the
simplest signature of a second order transition comes from its vanishing.
Also, the
strength of a first order transition is characterised in terms of
effective Lagrangians by the
coefficient of the term cubic in the scalar field.  This quantity is
greatly influenced once again by the corrected mass of the longitudinal
vector field [9].

An alternative procedure for avoiding infrared problems can be based on the
idea of dimensional reduction [10]. In this approach an effective
theory is derived for the static bosonic Matsubara modes, where
temperature dependent masses are created by non-static modes.

In a recent publication a saddle point approximation has been proposed
for the solution of the effective theory and tested on the example of the
pure scalar field theory [11].
The saddle point value of some
auxiliary variable serves for the effective mass-square of the scalar field.
The high-temperature phase corresponds to all of its components having
the same nonzero mass. Infrared problems are avoided.

A systematic expansion around the saddle point can be developed, where
the subsequent orders are organised in powers of the inverse number of the
components (large-N expansion).

Second order
transitions can be approached with this method only from the symmetric phase.
In case of first order transitions one can enter the broken phase down to the
temperature, where the symmetric phase becomes absolutely unstable.

A closely related approach has been put forward in Ref. [12], treating
the effective theory with cut-off(s). The cut-off value(s) is(are) chosen
proportional to the temperature in such a way that the final expression of
the effective potential should reproduce the leading high-T behaviour
determined in the 4D theory. This version of the reduced field theory has
been applied already to the SU(2) non-Abelian gauge theory, too [13].

In the present paper the strategy of Ref. [11] will be applied to the
Abelian Higgs-model. Integration over the non-static modes will be
performed at 1-loop level. In the potential piece of the effective
model terms up to quartic power in the static fields are retained
(higher powers being suppressed at high temperature). Both the scalar
field and the "thermal" component of the vector potential are treated
on equal footing. The dependence of the effective Lagrangian on the
static vector field is found by exploiting the 3D gauge invariance of
the effective theory. (Section 2)

The effective problem is solved by introducing auxiliary fields,
respectively conjugate to the scalar complex field
and the thermal component of the
vector field. The dynamics of these fields is treated in a
saddle point approximation. A simple argument shows that for any
nonzero value of $e$ the transition is of first order nature in our
approximation.
The saddle-point equations can be solved
analytically in the asymptotic weak coupling regime. (Section 3)

Finally, a numerical solution for small, but finite couplings is also
presented. Here the non-trivial question of the cancellation of 3-dimensional
infinities produced in the course of the derivation of effective models
is carefully investigated. Also detailed comparison with other approximate
treatments available in the recent litterature is given (Section 4).
Conclusions are summarized in Section 5.

\newpage
\noindent{\bf 2. Derivation of the effective model}
\bigskip
The field theory under investigation is defined by the following
Euclidean action
$$S=-\int d^{4}x\bigl [{1\over 4}F_{mn}F_{mn}+{1\over
2}|D_{m}\Phi |^{2}+{1\over 2}m^{2}|\Phi |^{2}+{\lambda\over 4!}
|\Phi |^{4}\bigr ],\eqno(2.1)$$
where $\Phi$ is a complex field and
$$F_{mn}=\partial_{m}A_{n}-\partial_{n}A_{m},~~~~~D_{m}\Phi
=(\partial_{m}+ieA_{m})\Phi,~~~~m=1,..,4.\eqno(2.2)$$
It is convenient to rewrite (2.1) in three-dimensional notation
(the "time"-components are indexed by $\tau$):
$$S=-\int d^{4}x\bigl [{1\over 4}F_{ij}F_{ij}+{1\over
2}(\partial_{\tau} A_{i})^{2}+{1\over
2}(\partial_{i}A_{\tau})^{2}-(\partial_{i}A_{i})(\partial_{\tau}A_{\tau})$$
$$~~~~~~~+{1\over 2}|\partial_{\tau}\Phi |^{2}+{1\over
2}|\partial_{i}\Phi |^{2}+{e^{2}\over 2}(A_{\tau}^{2}+A_{i}^{2})|\Phi
|^{2}+ {ieA_{\tau}\over
2}(\Phi\partial_{\tau}\Phi^{*}-\Phi^{*}\partial_{\tau} \Phi )$$
$$~~~~~~~+{ieA_{i}\over
2}(\Phi\partial_{i}\Phi^{*}-\Phi^{*}\partial_{i}\Phi )+{1\over
2}m^{2}|\Phi |^{2}+{\lambda\over 4!}|\Phi |^{4}\bigl ].\eqno(2.3)$$

The first step of the derivation is the separation of the static parts
of the fields:
$$A_{i}({\bf x},\tau )=A_{i}({\bf x})+a_{i}({\bf x},\tau ),~~~~
A_{\tau}({\bf x},\tau )=\rho ({\bf x})+a_{\tau}({\bf x},\tau ), ~~~~\Phi (
{\bf x},\tau )=\varphi_{0}({\bf x})+\varphi ({\bf x},\tau )
\eqno(2.4)$$
with the non-static parts fulfilling $\int a_{i}d\tau =\int
a_{\tau}d\tau =\int\varphi d\tau =0.$

The purely static part of the action is a three-dimensional theory of
the form
$$S^{(0)}=-\beta\int d^{3}x\bigl [{1\over 4}F_{ij}F_{ij}+{1\over
2}(\partial_{i}\rho )^{2}+{1\over
2}|\partial_{i}\varphi_{0}|^{2}+{e^{2}\over 2}(A_{i}^{2}+\rho^{2})
|\varphi_{0}|^{2}$$
$$~~~~~~~+{ieA_{i}\over
2}(\varphi_{0}\partial_{i}\varphi^{*}_{0}-\varphi_{0}^{*}
\partial_{i}\varphi_{0} )+{1\over
2}m^{2}|\varphi_{0}|^{2}+{\lambda\over 4!}|\varphi_{0} |^{4}\bigr
].\eqno(2.5)$$
The effective theory is obtained by integrating over the non-static
fields in the Gaussian approximation. For this operation it is the most
convenient to work in the static thermal gauge:
$$a_{\tau }=0.\eqno(2.6)$$
Since the three-dimensional (spatial) gauge-invariance is left intact, it
will be explicitly displayed by the resulting action.
\newpage
$$S^{(2)}_{ax}=\int d^{4}x\bigl [-{1\over 2}(\partial_{i}a_{j})^{2}+
{1\over 2}(\partial_{i}a_{i})^{2} -{1\over 2}(\partial_{\tau}a_{i})^{2}-
{1\over 2}|\partial_{\tau}\varphi |^{2}-{1\over 2}|\partial_{i}\varphi |^{2}$$
$$~~~~~~~-{1\over 2}e^{2}a_{i}^{2}|\varphi_{0}|^{2}-{1\over
2}e^{2}\rho^{2}|\varphi |^{2}-{ie\over 2}\rho
(\varphi\partial_{\tau}\varphi^{*}-\varphi^{*}\partial_{\tau}\varphi) $$
$$~~~~~~~-{1\over 2}m^{2}|\varphi |^{2}-{\lambda\over
4!}(\varphi^{2}_{0}\varphi^{*2}
+\varphi^{*2}_{0}\varphi^{2}+4|\varphi_{0}|^{2}|\varphi |^{2})$$
$$~~~~~~~-{ie\over
2}a_{i}(\varphi_{0}\partial_{i}\varphi^{*}+\varphi\partial_{i}
\varphi_{0}^{*}-\varphi^{*}_{0}\partial_{i}\varphi
-\varphi^{*}\partial_{i}\varphi_{0} )\bigr ]\eqno(2.7)$$
In (2.7) also $A_{i}({\bf x})$ is set to zero, since by the three-dimensional
gauge invariance this functional dependence of the effective action can
be restored from the kinetic part of the action. In this paper we
neglect the "renormalisation" of the kinetic part, only the effect of
the non-static modes on the potential will be discussed. (The
modification of the kinetic term can be studied using the procedure
described in Ref.[14]). This is the reason why we use only the
$\varphi_{0}$ and $\rho$ background fields, which are set for our
restricted purpose to real constants.

The most convenient form of (2.7) for the functional integration is
found after several integrations by parts and by exploiting the
periodicity of the small fields in $\tau$:
$$S^{(2)}_{ax}=-\int d^{4}x\bigl\{{1\over 2}a_{i}\bigl
[(-D_{E}+e^{2}\varphi^{2}_{0} )\delta_{ij}+\partial_{i}\partial_{j}\bigr
]a_{j} $$
$$~~~~~~~~~~+{1\over
2}\varphi_{1}(-D_{E}+e^{2}\rho^{2}+m^{2}+{\lambda\over
2}\varphi^{2}_{0})\varphi_{1} $$
$$~~~~~~~~~~+{1\over
2}\varphi_{2}(-D_{E}+e^{2}\rho^{2}+m^{2}+
{\lambda\over 6}\varphi^{2}_{0})\varphi_{2}$$
$$~~~~~~~~~+e\varphi_{0}\varphi_{2}\partial_{i}a_{i}+e\rho
(\varphi_{2}\partial_{\tau}
\varphi_{1}-\varphi_{1}\partial_{\tau}\varphi_{2}) \bigr\},\eqno(2.8)$$
where the notations $D_{E}=\partial^{2}_{i}+\partial^{2}_{\tau}$ and
$\varphi =\varphi_{1}+i\varphi_{2}$ are introduced. The result of the
integration can be expressed in familiar Fourier-sums:
$$\beta V~U_{eff}(\rho ,\varphi_{0})=\beta V~\bigl [{1\over
2}m^{2}\varphi_{0}^{2}+{\lambda\over 4!}\varphi_{0}^{4}+{e^{2}\over
2}\rho^{2} \varphi_{0}^{2}\bigr ]$$
$$~~~~~~~~+\sum_{n\neq 0}\sum_{k}\ln (k^{2}+e^{2}\varphi^{2}_{0})+
{1\over 2}\sum_{n\neq 0}\sum_{k}\ln
(\omega_{n}^{2}k^{4}+A_{1}\omega^{2}_{n}k^{2}+A_{2}\omega^{4}_{n}
+B_{1}\omega_{n}^{2}+B_{2}k^{2}+C),\eqno(2.9)$$
with $k^{2}=\omega^{2}_{n}+{\bf k}^{2}, \omega_{n}=2\pi\beta^{-1}n$ and
\newpage
$$A_{1}=2e^{2}\rho^{2}+2m^{2}+({2\lambda\over 3}+e^{2})\varphi_{0}^{2},~~~~
A_{2}=-4e^{2}\rho^{2},$$
$$~~~~~~~B_{1}=(e^{2}\rho^{2}+m^{2}+{\lambda\over
2}\varphi_{0}^{2})(e^{2}\rho^{2} +m^{2}+({\lambda\over
6}+e^{2})\varphi_{0}^{2})- 4e^{4}\rho^{2}\varphi_{0}^{2},$$
$$~~~~~~~B_{2}=e^{2}\varphi^{2}_{0}(e^{2}\rho^{2}+m^{2}+{\lambda\over
6}\varphi_{0}^{2}),~~~~
C=e^{2}\varphi_{0}^{2}(e^{2}\rho^{2}+m^{2}+{\lambda\over
2}\varphi_{0}^{2})(e^{2}\rho^{2}+m^{2}+{\lambda\over 6}\varphi_{0}^{2}).
\eqno(2.10)$$

For the high-$T$ asymptotics the logarithms in (2.9) are expanded into
power series, which can be stopped at $o(k^{-6},\omega^{-2}k^{-4},...)$
since these terms would contribute $o(m^{2}/T^{2})$ to the
potential. The evaluation of the $n\neq 0$ sums is followed by that of
the $\bf k$-integrals $(V\rightarrow\infty$) with a sharp cut-off
$\Lambda$. Collecting the contributions up to quartic terms one finds

$$U_{eff}={1\over 2}\varphi_{0}^{2}\bigl [m^{2}+({2\lambda\over
3}+3e^{2 })({\Lambda^{2}\over 8\pi^{2}}-{\Lambda\over
2\pi^{2}\beta}+{1\over 12\beta^{2}})+{m^{2}\over
8\pi^{2}}({2\lambda\over 3}+e^{2})- {m^{2}\over
4\pi^{2}}({2\lambda\over 3}+3e^{2})I\bigr ]$$
$$~~~~~~~~+{1\over 2}\rho^{2}[ e^{2}({1\over 3\beta^{2}}+{m^{2}\over
4\pi^{2}})-{\Lambda e^{2}\over \pi^{2}\beta}]+{e^{4}\rho^{4}\over 24\pi^{2}}$$
$$~~~~~~~~+{1\over 2}e^{2}\rho^{2}\varphi_{0}^{2}[1+({2\lambda\over
3}+5e^{2}){1\over 8\pi^{2}}-{3e^{2}\over 2\pi^{2}}I]$$
$$~~~~~~~~+{\varphi^{4}_{0}\over 4!}[\lambda+3({5\over
36}\lambda^{2}+{1\over 2}e^{2}\lambda +{3\over 2}e^{4})({1\over 2\pi^{2}}-I)]
.\eqno(2.11)$$
The abbreviation $I$ denotes the logarithmically divergent integral
$$I=\int_{0}^{\Lambda\beta}dx({1\over 2x}+{1\over x(e^{x}-1)}-{1\over
x^{2}}).\eqno(2.12) $$

The renormalisation prescriptions for the couplings are chosen by fixing the
second derivative with respect to the scalar field to the renormalized mass,
 the fourth derivative to the renormalised scalar self-coupling and
the coefficient of the scalar - thermal vector component
vertex to the electric charge in the effective potential at $T=0$:
$$m_{R}^{2}=m^{2}+({2\lambda\over 3}+3e^{2}){\Lambda^{2}\over
8\pi^{2}}-{m^{2}\over 4\pi^{2}}({2\lambda\over 3}+3e^{2})I+{m^{2}\over
8\pi^{2}} ({2\lambda\over 3}+e^{2}),$$
$$e_{R}^{2}=e^{2}\bigl [1+({2\lambda\over 3}+5e^{2}){1\over
8\pi^{2}}-{3e^{2}\over 2\pi^{2}}I\bigr ],$$
$$\lambda_{R}=\lambda +{3\over \pi^{2}}({5\over 36}\lambda^{2}+{1\over
6}e^{2}\lambda +{3\over 2}e^{4})({1\over 2}-I).\eqno(2.13)$$
The renormalised effective potential takes a form, where the
three-dimensional "counter\-terms" are displayed explicitly
(below we omit the subscript $R$ from $e$ and $\lambda$):
\newpage
$$U_{eff,R}={1\over 2}\varphi_{0}^{2}\bigl [m^{2}_{R}+({2\lambda\over
3}+3e^{2}){1\over 12\beta^{2}}-({2\lambda\over 3}+3e^{2}){\Lambda\over
2\pi^{2}\beta}\bigr ] $$
$$~~~~~~~~+{1\over 2}e^{2}\rho^{2}\phi^{2}_{0}+{1\over
2}\rho^{2}[e^{2}({1\over 3\beta^{2}}+{m^{2}\over 4\pi^{2}})-{\Lambda
e^{2}\over \pi^{2}\beta}]+{e^{4}\rho^{4}\over
24\pi^{2}}+{\lambda\varphi_{0}^{4}\over 4!}.\eqno(2.14)$$
It is notable that by the gauge invariance of the full theory no
quadratic divergence has appeared related to the $\rho$-field. However,
in the effective model, where it plays the role of an extra scalar
field, the necessity of an appropriate $\rho$-mass renormalisation is
indicated by (2.14). Also the plasmon mass in front of $\rho^{2}$ is
correctly reproduced.

In order to work with truly three-dimensional quantities the following
rescaling is done:
$$U_{i}=\sqrt{\beta}A_{i},~~~r=\sqrt{\beta}\rho,~~~
\phi=\sqrt{\beta}\varphi_{0},$$
$$g={\lambda\over \beta},~~~~\epsilon ={e\over \sqrt{\beta}}.\eqno(2.15)$$
With this notation the final form of the 3D effective theory is written
as
$$S_{3D}=-\int d^{3}x\bigl [{1\over 2}\bigl
((\partial_{i}U_{j})^{2}-(\partial_{i}U_{i})^{2}\bigr)+{1\over
2}|(\partial_{i}+i\epsilon U_{i})\phi|^{2}$$
$$~~~~~~~~+{1\over 2}m_{\phi}^{2}|\phi |^{2}+{1\over
2}(\partial_{i}r)^{2}+{1\over 2}m_{r}^{2}r^{2}+{g\over 4!}|\phi |^{4}$$
$$~~~~~~~~+{\epsilon^{4}\beta\over 24\pi^{2}}r^{4}+{1\over
2}\epsilon^{2}r^{2}|\phi|^{2}\bigr ],\eqno(2.16)$$
where the $U_{i} (A_{i})$-dependence is restored and the abbreviations
$$m_{\phi}^{2}=m_{R}^{2}+({2\lambda\over 3}+3e^{2}){1\over
12\beta^{2}}-({2\lambda\over 3}+3e^{2}){\Lambda\over 2\pi^{2}\beta},$$
$$m_{r}^{2}=e^{2}({1\over 3\beta^{2}}+{m^{2}\over 4\pi^{2}})-{\Lambda
e^{2}\over \pi^{2}\beta}\eqno(2.17)$$
are introduced.

Eq. (2.16) differs slightly
from the effective theory of Ref.[12]. It implies the
complete determination of the $r$-potential. Also the correct
$T$-dependence of $m_{\phi}^{2}$ and $m_{r}^{2}$ is achieved
without the somewhat artificial step of introducing different
3D momentum cut-off scales
 for different fields.
The investigation below concentrates on the suppression of the
Higgs-effect with increasing temperature. Therefore, in the next
section only the spontanous creation of a vacuum expectation value for
$\phi (x)$ will be investigated. However, in some physically more
appealing situations, one might attempt a broader unconstrained analysis
of the competing condensations of $\phi$ and $r$ in different regions of
the $(\lambda - e^{2})$-plane.
\vfill\eject
\noindent{\bf 3. Two-variable saddle point approximation of the
effective potential}
\bigskip
The approximate solution we are going to discuss in this section shares
with other improved perturbative methods the feature of allowing some
masses to take temperature dependent values. Instead of generating
these expressions from summing infrared singular diagrams or from
self-consistent equations for polarisation functions, we extend the
model by introducing auxiliary variables, and include the contribution from
their saddle-point values to the masses. The actual procedure will
influence the effective masses of the scalar and the longitudinal vector
components.

The auxiliary fields are introduced by the standard
Hubbard\- -Stratonovich transformation [15] of the quartic terms in the action
(2.16):
$$\int_{C-i\infty}^{C+i\infty}d\chi\exp\bigl\{{1\over
8}\chi_{\alpha}B_{\alpha\beta} \chi_{\beta}-{1\over
2}\chi_{\alpha}A_{\alpha} \bigr\}={\rm const.}\times\exp\bigl\{ -{1\over 2}
A_{\alpha} B_{\alpha\beta}^{-1}A_{\beta}\bigr\}\eqno(3.1)$$
($\alpha =1,2$). In the present case the matrix $B$ and the vector $A$
are found from (2.16)
$$A=(r^{2},~|\phi|^{2}),~~~B={4\over \epsilon^{4}}({\beta g\over
36\pi^{2}} -1)^{-1}\left(\matrix{{g\over 12}&-{\epsilon^{2}\over 2}\cr
-{\epsilon^{2}\over 2}&{\epsilon^{4}\beta\over
12\pi^{2}}\cr}\right)\eqno(3.2)$$

The extended form of the 3D action reads
$$S_{3D}[\chi_{\alpha},\phi ,U_{i},r]=-\int d^{3}x\bigl\{{1\over 2}
[(\partial_{i}U_{j})^{2}-(\partial_{i}U_{i})^{2}]+{1\over
2}|(\partial_{i}+i\epsilon U_{i})\phi|^{2}$$
$$~~~~~~~~+{1\over 2}M_{\phi}^{2}|\phi |^{2}+{1\over
2}M_{r}^{2}r^{2} +{1\over 2}(\partial_{i}r)^{2}$$
$$~~~~~~~~-{1\over 2\epsilon^{4}}({\beta g\over
36\pi^{2}}-1)^{-1}({g\over
12}\chi_{1}^{2}-\epsilon^{2}\chi_{1}\chi_{2}+{\epsilon^{4}\beta\over
12\pi^{2}}\chi_{2}^{2}) \bigr\}\eqno{(3.3)}$$
with the mass squares modified by the auxiliary fields:
$$M_{\phi}^{2}=m_{\phi}^{2}+\chi_{2},~~~M_{r}^{2}=m_{r}^{2}+\chi_{1}.
\eqno{(3.4)}$$

We approximate the $\chi_{\alpha}$-integrals by real, optimised
saddle-point values and evaluate the $\phi ,U_{i}$ and $r$ integrals on
their background. This is the standard procedure when the number of
components of the scalar field is large. Here the expected nature of the
$T$-dependent corrections to the $\phi$- and $r$-masses suggests this approach.
There is no formal argument for fast convergence of the expansion around this
saddle. An eventual agreement with results of the improved perturbation
theory could be considered as a positive sign for the adequacy of our
method.
\vfill\eject
For the 1-loop evaluation of the effective 3D potential the value of
the the $\phi$-field is shifted by $\bar\phi$. The integration over the
variables $U_{i}$ and $\phi$ are decoupled with an appropriate gauge-fixing
 function added to (3.3) [14]:
$$S_{gf}={1\over 2\alpha}\int
d^{3}x(\partial_{i}U_{i}+\alpha\epsilon\bar\phi \phi_{2})^{2}.\eqno{(3.5)}$$
The corresponding Faddeev-Popov ghost contribution is
$$U^{3D}_{eff,ghost}={\rm Tr}\log
(-\partial_{i}^{2}+\alpha\epsilon^{2}\bar\phi^{2} ).\eqno{(3.6)}$$

The formal result of the Gaussian functional integrations is summarised
as
$$VU_{eff}^{(3D)}={1\over 2}{\rm Tr}\log \bigl
[\delta_{ij}(-\partial_{k}^{2}+\epsilon^{2} \bar\phi^{2})+(1-{1\over
\alpha})\partial_{i}\partial_{j}\bigr ]-{\rm
Tr}\log (-\partial_{k}^{2}+\alpha\epsilon^{2} \bar\phi^{2})$$
$$~~~~~~+{1\over 2}{\rm Tr}\log (-\partial_{k}^{2}+M_{r}^{2})+{1\over
2}{\rm Tr}\log (-\partial_{k}^{2}+M_{\phi}^{2})+{1\over 2}{\rm
Tr}\log
(-\partial_{k}^{2}+M_{\phi}^{2}+\alpha\epsilon^{2}\bar\phi^{2} )$$
$$~~~~~~+V\bigl [{1\over 2}M_{\phi}^{2}\bar\phi^{2}+{1\over
2\epsilon^{4}} (1-{\beta g\over 36\pi^{2}})^{-1}({g \over
12}\chi_{1}^{2} +{\epsilon^{4}\beta\over 12\pi^{2}}\chi_{2}^{2}-
\epsilon^{2}\chi_{1}\chi_{2} )\bigr ].\eqno{(3.7)}$$
The traces in the above expression were evaluated with sharp cut-off
$\Lambda_{3}$:
$$U_{eff}^{(3D)}=\Lambda_{3}{1\over
4\pi^{2}}(2\epsilon^{2}\bar\phi^{2}+M_{r}^{2} +2M_{\phi}^{2})+{1\over
2}M_{\phi}^{2}\bar\phi^{2}$$
$$~~~~~~~~+{1\over 2\epsilon^{4}}(1-{\beta g\over
36\pi^{2}})^{-1} ({g\over 12}\chi^{2}_{1}+{\epsilon^{4}\beta\over
12\pi^{2}}\chi_{2}^{2}-\epsilon^{2}\chi_{1}\chi_{2}) $$
$$~~~~~~~~-{1\over
12\pi}[-(\alpha\epsilon^{2}\bar\phi^{2})^{3/2}+2\epsilon^{3}|\bar\phi
|^{3}+M_{r}^{3}+M_{\phi}^{3}+(M_{\phi}^{2}+
\alpha\epsilon^{2}\bar\phi^{2})^{3/2}].\eqno{(3.8)}$$

The finiteness of the expression in  the third line of (3.8) requires
the renormalisation of the $\chi_{\alpha}$ values [11] as seen from the
detailed
expressions of $M_{\phi}^{2}$ and $M_{r}^{2}$:
$$M_{\phi}^{2}=m_{\phi ,R}^{2}-({2\lambda\over 3}+3e^{2}){\Lambda\over
2\pi^{2}\beta}+\chi_{2} ,$$
$$M_{r}^{2}=m_{r,R}^{2}
-{\Lambda e^{2}\over \pi^{2}\beta}+\chi_{1},$$
$$m_{\phi ,R}^{2}=m_{R}^{2}+({2\lambda\over 3}+3e^{2}){1\over
 12\beta^{2}},~~~~~m_{r,R}^{2}=e^{2}({1\over 3\beta^{2}}+
{m_{R}^{2}\over 4\pi^{2}}).\eqno{(3.9)}$$
The cancellation of the infinities implies the existence of the cut-off
independent saddle-points
$$\chi_{1R}=\chi_{1}-{\Lambda e^{2}\over
\pi^{2}\beta},~~~~\chi_{2R}=\chi_{2}- ({2\lambda\over
3}+3e^{2}){\Lambda\over 2\pi^{2}\beta}.\eqno{(3.10)}$$

One reexpresses the $\chi$-dependent part of (3.8)  and throws away the
(infinite) constants arising in this way. Below, we write down in
separate expressions the linearly divergent and the finite parts  of the
effective potential after reintroducing 4D notations. For the
simplification of the expressions from this point we use the $\alpha
=0$ gauge.
$$U_{eff,div}=\Lambda_{3}{1\over
4\pi^{2}}(2e^{2}\bar\Phi^{2}+\chi_{1R}+2\chi_{2R} )$$
$$~~~~-\Lambda{1\over 2e^{2}}(1-{\lambda\over
36\pi^{2}})^{-1}[\chi_{1R}{1\over 2\pi^{2}}({\lambda \over
3}+3e^{2}) -\chi_{2R}{e^{2}\over \pi^{2}}({1\over
12\pi^{2}}({2\lambda\over 3}+3e^{2})-1)],\eqno(3.11)$$

$$U_{eff,finite}={1\over 2}[m_{R}^{2}+({2\lambda\over
3}+3e^{2}){T^{2}\over 12}+\chi_{2R}]\bar\Phi^{2}$$
$$~~~~+{1\over 2e^{4}}(1-{\lambda\over 36\pi^{2}})^{-1}({\lambda\over
12}\chi_{1R}^{2}+ {e^{4}\over
12\pi^{2}}\chi_{2R}^{2}-e^{2}\chi_{1R}\chi_{2R})$$
$$-{T\over 12\pi}[2e^{3}\bar\Phi^{3}+(e^{2}({T^{2}\over
3}+{m_{R}^{2}\over
4\pi^{2}})+\chi_{1R})^{3/2}+2(m_{R}^{2}+({2\lambda\over
3}+3e^{2}){T^{2}\over 12}+\chi_{2R})^{3/2}].\eqno(3.12)$$

The vanishing of (3.11) should be automatic, when
$\chi_{\alpha,R}(\bar\Phi )$ extremising (3.12) is substituted into it.
The only freedom we have is to choose $\Lambda_{3}$ in constant
proportion to $\Lambda$. A sufficient condition for this is to find
$$\chi_{\alpha R}=c_{\alpha 1}+c_{\alpha 2}\bar\Phi^{2},\eqno(3.13)$$
since the field-independent parts of $U_{eff,div}$ can be omitted. The
fulfillment of (3.13) should be checked for any solution to be
presented below.

The saddle point coordinates should be determined from the following
extremal conditions:
$${\partial U_{eff,finite}\over \partial\chi_{1R}}=-{T\over
8\pi}[m_{r,R}^{2}+\chi_{1R}]^{1/2}+{1\over 2e^{4}}(1-{\lambda\over
36\pi^{2}})^{-1}({\lambda\over 6}\chi_{1R}-e^{2}\chi_{2R})=0,$$
$${\partial U_{eff,finite}\over \partial\chi_{2R}}={1\over
2}\bar\Phi^{2}+{1\over 2e^{2}}(1-{\lambda\over
36\pi^{2}})^{-1}({e^{2}\over 6\pi^{2}}\chi_{2R}-\chi_{1R})-
{T\over 4\pi}[m_{\phi ,R}^{2}+\chi_{2R}]^{1/2}=0.\eqno(3.14)$$

The location of the non-trivial minima of the effective potential is
determined by the equation
$${\partial U_{eff,finite}\over \partial\bar\Phi^{2}}={1\over
2}[m_{\phi ,R}^{2}+\chi_{2R}(\bar\Phi )]-{e^{3}T\over 4\pi}\bar\Phi =0.
\eqno(3.15)$$
\vfill\eject
{\it The phase structure in the $\lambda -e^{2}$ plane}
\bigskip
It is not difficult to argue that eqs.(3.14-15) imply a first order
transition for any non-zero value of $e^{2}$.

{}From the first equality in eq.(3.14) for $e^{2}\rightarrow 0$ one
realizes that $\chi_{1R}$ and $\chi_{2R}$ are related to each other for
all values of $\bar\Phi$:
$${\lambda\over 6}\chi_{1R}\approx e^{2}\chi_{2R}.\eqno(3.16)$$
Using this relation in the second equality one solves it for the full
"effective" mass
$$H^{1/2}=\bigl [\chi_{2R}+m_{\phi ,R}^{2}\bigr ]^{1/2}\eqno(3.17)$$
leading to
$$H^{1/2}={1\over 2}\bigl [-{\lambda T\over 12\pi}+\bigl
({\lambda^{2}T^{2}\over (12\pi )^{2}}+4({\lambda\over
6}\bar\Phi^{2}+m_{\phi ,R}^{2})\bigr )^{1/2}\bigr ].\eqno(3.18)$$
Defining the temperature $T_{2}$, where the symmetric minimum becomes
absolutely unstable ($H(\bar\Phi =0)=0$), a simple expression is derived
for $H(\bar\Phi ,T_{2})$:
$$H(\bar\Phi )^{1/2}={1\over 2}[-{\lambda T_{2}\over
12\pi}+({\lambda^{2}T^{2}_{2}\over (12\pi )^{2}}+{2\lambda\over
3}\bar\Phi^{2})^{1/2} ].\eqno(3.19)$$
After substituting (3.19) into (3.15) one sees the origin transformed
into local maximum, while a non-trivial minimum is found at
$$\bar\Phi_{min}={eT_{2}\over 2\pi}+o(e^{2}T_{2}).\eqno(3.20)$$

This behaviour has been confirmed by the numerical solution of (3.14-15)
for $\lambda =10^{-5}$. It is interesting to note that Fig. 1 also
illustrates the difficulty of such test. The solution of (3.14) with
restricted accuracy leads to an apparent stopping of the position of
$\Phi_{min}$ when the charge $e$ becomes small enough. With gradually
improved accuracy one observes the formation of a linear envelope
in the interval $e\in (10^{-7}, 10^{-3})$.

The form of $U_{eff}(T_{2})$ shows that for $T_{c}>T_{2}$ the distance
between the origin and the non-trivial minimum is finite,
the transition is of first order nature. However, the effect of
contributions from fluctuations around the saddle point might
change this conclusion.
\bigskip

{\it The asymptotic weak coupling solution}
\bigskip
A fully analytic selfconsistent construction will be presented
for the solution of the gap equations when $\lambda << 1, e^{2} << 1$.
Special attention will be paid to the cancellation of the linear divergencies.

It is convenient to introduce the notations
$$m^{2}_{eff}=m^{2}_{\phi ,R}-{e^{2}Tm_{r,R}\over 4\pi}$$
and
$$Q={\lambda\over 6}-{e^{4}T\over 8\pi m_{r,R}}(1-{\lambda\over 36\pi^{2}}
).\eqno{(3.21)}$$
The temperature range covered by our analysis will be controlled by
requiring
$$m_{eff}^{2}\sim o(e^{n}T^{2}).\eqno{(3.22)}$$
{}From the previous subsection it is clear, that $T_{c}^{2}>T_{2}^{2}\approx
-36m_{R}^{2}/(2\lambda +9e^{2})$. Therefore, $m_{r,R}\sim o(eT)$.

Our fundamental assumption is that in the relevant $T$-interval $\chi_{1}<<
m^{2}_{r,R}$, which will be checked on the solution. Under this assumption
 we expand the square-root of the first equation of (3.14),
and find an approximate linear
relationship between $\chi_{1}$ and $\chi_{2}$. Using it in the second equation
of (3.14) one writes for $m_{\phi ,R}^{2}+\chi_{2R}$:
$$m_{\phi ,R}^{2}+\chi_{2R}+{QT\over 2\pi}\sqrt{m_{\phi ,R}^{2}+\chi_{2R}}
-(Q\bar\Phi^{2}+m_{eff}^{2})+~o\bigl({e^{2}T\chi_{1R}\over m^{3}_{r,R}},
{e^{4}T\chi_{2R}\over m_{r,R}}\bigr )=0.\eqno{(3.23)}$$

Its solution yields
$$\chi_{1R}=-{\lambda e^{4}Tm_{r,R}\over 144\pi^{3}Q}
+{Qe^{2}T^{2}\over 8\pi^{2}}
+e^{2}\bar\Phi^{2}$$
$$~~~~~~~~~~~-{e^{2}T\over 2\pi}\bigl [{Q^{2}T{2}\over
16\pi^{2}}+Q\bar\phi^{2}+m_{eff}^{2}\bigr ]^{1/2}+
o\bigl({e^{4}T\chi_{1R}\over Qm_{r,R}^{3}},{e^{6}T
\chi_{2R}\over Qm_{r,R}}\bigr),$$
$$\chi_{2R}=-{e^{2}Tm_{r,R}\over 4\pi}+{Q^{2}T^{2}\over 8\pi^{2}}
+Q\bar\Phi^{2}$$
$$~~~~~~~~~~~-{QT\over 2\pi}\bigl [{Q^{2}T^{2}\over 16\pi^{2}}
+Q\bar\Phi^{2}+m_{eff}^{2}\bigr ]^{1/2}
+o\bigl({e^{2}T\chi_{1R}^{2}\over m_{r,R}^{2}},{e^{4}T\chi_{2R}
\over m_{r,R}}\bigr).\eqno{(3.24)}$$

The range of $\bar\Phi$, where this solution is required to be
valid should cover the anticipated second non-trivial minimum
of the potential. The order of magnitude of its distance
from the origin can be estimated with help of (3.15), when the
 above expression of $\chi_{2}$ is substituted:
$$m_{eff}^{2}+Q\bar\Phi^{2}_{min}-({eT\over 2\pi})^{3/2}{Q\over 2}
\bar\Phi_{min}^{1/2}-{e^{3}T\over 2\pi}\bar\Phi_{min}=0.\eqno(3.25)$$

That part of the $\lambda -e^{2}$-plane will be considered,
where terms proportional to $Q$ can be neglected in the equation of
$\bar\Phi_{min}$:
$$\bar\Phi_{min}={2\pi m_{eff}^{2}\over e^{3}T}\sim~o(e^{n-3}T).
\eqno(3.26)$$
Since the requirement $\chi_{1R}<<m_{r,R}^{2}$  implies $\bar\Phi<<T$,
one has to require $n>3$. {\it A posteriori} checking the order of
magnitude of the omitted terms in (3.25), the consistency gives the condition
$$Q<o(e^{m}),~~~~m=max(6-n,{n\over 2}).\eqno{(3.27)}$$
This estimate can be used for further simplifying the expressions of the
saddle point values:
$$\chi_{1R}=-{e^{2}T\over 2\pi}m_{eff}+e^{2}\bar\Phi^{2}+o(Q^{2}T^{2}),$$
$$\chi_{2R}=-{e^{2}T\over 4\pi}m_{r,R}-{QT\over 2\pi}m_{eff}+Q\bar\Phi^{2}
+o(Q^{2}T^{2}).\eqno(3.28)$$

The order of magnitude estimate of the errors in (3.28) allows for $n$ the
range
$3<n<4$. It is this range which fixes, how close we can tune our solution to
the
$m_{eff}=0$ temperature. (It is worth to notice that from here by the
definition of $Q$ also $\lambda\sim o(e^{3})$ follows. Also one explicitly sees
that $\chi_{1R}$ is negligible relative to $m_{r,R}^{2}$.)

Before going to the description of the behaviour of $U_{eff}$, we consider
(3.11), the linearly divergent piece of the potential. One finds that
choosing
 $$\Lambda_{3}\approx\Lambda (1+{\lambda\over 9e^{2}})\eqno(3.29)$$
 its cancellation is ensured, since our solution (3.28) is conform with (3.13).

The leading expression of the potential then is given by
$$U_{eff}={1\over 2}m_{eff}^{2}\bar\Phi^{2}-{T\over 6\pi}e^{3}\bar\Phi^{3}
+{Q\over 4}\bar\Phi^{4}.\eqno(3.30)$$
Up to the modification $\lambda /6\rightarrow Q$ this expression coincides with
the improved form derived by Arnold [6]. It has been checked on the numerical
 solution ($\lambda =10^{-5}, e^{2}=10^{-4}$) that the solution is very well
represented by the above formulae. Note, however, that the only reasonable
comparison was to calculate the effective potentials at the transition
temperatures of the respective approximations.

The weakening of the strength of the
first order transition for small $e$ can be characterised by the scaling
of the height of the barrier between the degenerate minima at $T_{c}$.
{}From (3.30) one finds the power $e^{3n-6}$. In view of the allowed range of
$n$
this leads to a "height-exponent" between 3 and 6.
\newpage
\noindent{\bf 4. Small coupling numerical solution}
\bigskip
The solution of the system (3.14) exists whenever the squre roots
appearing in these equations take real values. We have explicitly
found $\chi_{\alpha ,R}$ in the full $\bar\Phi$ interval covering both minima
of
the potential for the coupling space $10^{-7}\leq \lambda\leq 10^{-1},
10^{-6}\leq e^{2}\leq 0.5$.
 In the whole region the symmetry restoration proceeds through first
order transition.
The situation is well illustrated by Fig.2, where also curves of the
effective potential below and above $T_{c}$ are displayed.
All quantities are displayed in proportion to an appropriate power of
the $T=0$, tree level expectation value of $\bar\Phi$.
The complex nature of the effective potential expected in
intervals of non-convexity does not show up in the present approximation.

Our solution for these not asymptotically small couplings
$(\lambda =0.07, e^{2}=.32)$ clearly deviates
from the weak coupling solution of the previous section.
The measure of the deviation can be quantitatively assessed by Fig.3, where
the numerically obtained $\chi_{2R}$ is displayed together with
(3.28). The deviation is small near $\bar\Phi =0$, but
 increases gradually towards larger $\bar\Phi$.
It is notable that $\chi_{2}$ yields a negative contribution
 to the effective mass of $\bar\Phi$ near the origin. This represents a
tendency
 towards the instability of a homogenous background expected  for small
values of the field and for temperatures below the appearance of the
non-trivial metastable local minimum.
In case of $\chi_{1R}$ the deviation of the numerical solution
from the asymptotic weak coupling formulae cannot be resolved visually
on the same scale.

{}From the point of view of selfconsistency the cancellation of the
linearly divergent piece should be investigated. Any deviation
from zero would point to the importance of corrections to the
saddle point approximation. In Fig.4 the coefficient of $\Lambda_{3}$
in eq.(3.11) is displayed ($\Gamma$) against the same quantity in
front of $\Lambda (\Delta$). The observed fully linear behaviour
makes our leading approximation selfconsistent in an extended
part of the $(\lambda ,e^{2})$ plane. The slope is found extremely
 close to unity, supporting the validity of the weak coupling relation (3.20).

Our results can be compared quantitatively with curves of the effective
potential published in [7] for not asymptotically small couplings.
The values $\lambda =0.15, e^{2}=0.25$ correspond to Fig.8 of Ref [7].
The temperature closest to the actual transition, which can be faithfully
described perturbatively according to the authors is $T^{*}=T_{V}$ (in
their notation), the temperature where the infrared sensitivity of the vector
contribution to the potential becomes important. Its value for the above
coupling point is $T_{V}=0.647$. Since there the deviation of
$U_{eff}(0,T^{*})$ from $U_{eff}(\Phi_{min},T^{*})$ is at the $10^{-10}$
level, $T^{*}$ for all practical purposes can be considered as $T_{c}$ of
the approximation scheme [7] in this point. Our method leads for the
same couplings to $T_{c}=0.646$. The barrier height between degenerate minima
 at $T_{c}$ in proportion of the fourth power of the $T=0$, classical vacuum
expectation value of the Higgs field in the respective approximations
is $U_{eff}(\bar\Phi_{max},T_{c})=1.76\times 10^{-5} [7],~ 1.48\times 10^{-5}$
(present work). Finally, the Higgs expectation value at the transition in
proportion to its $T=0$, classical value is $\bar\Phi_{min}(T_{c})=0.492~ [7],
{}~0.463$ (present work).

\newpage
\noindent{\bf 5. Conclusions}
\bigskip
In this paper we have described in detail a two-variable saddle-point
improvement of the 1-loop finite temperature effective potential
of the Abelian Higgs-model. Its
main advantage is that it avoids the need to discuss explicitly the
delicate problem of infrared sensitivity. On the other hand it is quite
complicated to go beyond the leading saddle-point approximation in
the Abelian case compared to the pure scalar model with $N=\infty$
components.

Nevertheless the practical renormalisability of the effective potential
beyond asymptotically small couplings
 allows a rather clean interpretation of the leading order results beyond
the domain of its strict applicability. The good quantitative agreement
with more standard perturbative improvements gives confidence in applying
the method to more relevant systems, too.  Explicit calculations of the
$o(e^{2},\lambda^{2})$ corrections in the scheme of [7], and the evaluation
of the fluctuations around the saddle point in the present approximation
might decide if this spectacular agreement is more than just an accident.

The first order transition we observe is extremely weak. The barrier
between the degenerate minima at the transition seems to scale like
$\sim e^{3-6}$.
\bigskip
\noindent{\bf Acknowledgements}
\bigskip
The authors thank W. Buchm\"uller for kindly providing some data for
comparing the present method with his selfconsistent improved perturbative
approach. Also remarks by F. Karsch and J. Polonyi are gratefully acknowledged.
The research of A.P. was partly supported by
the EC-grant ERB3510PL920739.
\bigskip
\noindent{\bf Figure Captions}
\bigskip
\noindent
{\bf  Fig.1} Shifting of the position of the non-trivial minimum of the
effective potential at $T_{2}$ with
decreasing $e$ as found from the numerical solution. The leveling off
is the result of the finite accuracy of the solution. The true behavior
 as it is predicted by the weak coupling analysis
is given by the common envelope.\par
\noindent
{\bf Fig.2} The effective potential for $\lambda =0.07$ and $e^{2}=0.32$.
The three curves illustrate the ability of our method to describe the
neighbourhood of the transition in both phases\par\noindent
{\bf Fig.3} Comparison of the weak coupling and the numerical solutions of
$\chi_{2R}$ in the same $\lambda, e^{2}$ point as for Fig.2 ($T=T_{c})$.
The curve 1 gives the numerical solution for $\chi_{2}$, curve 2
represents its weak coupling solution (3.28).\par\noindent
{\bf Fig.4} Illustration of the linear relation between $\Lambda_{3}$ and
$\Lambda$ for couplings as above. The coefficient of $\Lambda_{3}$
is denoted by  $\Gamma$, that of $\Lambda$ by $\Delta$ ($T=T_{c}$),

\newpage
\noindent {\bf References}
\item{1.}A.G. Cohen, D.B. Kaplan and A.E.Nelson
Progress in Electroweak Baryogenesis, UCSD-PTH-93-02, BUHEP-93-4
\item{2.}V.A. Kuzmin, V.A. Rubakov and M.E. Shaposhnikov,
Phys. Lett. {\bf 155B} (1985) 36
\item{3.} A.D. Sakharov, JETP Letters {\bf 5} (1967) 24
\item{4.} L. McLerran, Phys. Rev. Lett. {\bf 62} (1989) 1075
\item{5.} L. Dolan and R. Jackiw, Phys. Rev. {\bf D9} (1974) 3320\\
    S. Weinberg, Phys. Rev. {\bf D9} (1974) 3357
\item{6.} M.E. Carrington, Phys. Rev. {\bf D45} (1992) 2933\\
        P. Arnold, Phys. Rev. {\bf D46} (1992) 2628
\item{7.} W. Buchm\"uller, T. Helbig and D. Walliser,
First Order Transitions in Scalar Electrodynamics, DESY-92-151, and
DESY-93-021
\item{8.} J.R. Espinosa, M. Quir\'os and F. Zwirner, Phys. Lett. {\bf 191B}
(1992) 115 and CERN-TH.6577/92
\item{9.} M. Dine, R.L. Leigh, P. Huet, A. Linde and D. Linde,
Phys. Rev. {\bf D46} (1992) 550
\item{10.} T. Appelquist and R. Pisarski, Phys. Rev {\bf D23} (1981) 2305\\
        N.P. Landsman, Nucl. Phys. {\bf B322} (1989) 498
\item{11.} H. Meyer-Ortmanns and A. Patk\'os, Phys. Lett. {\bf 297B} (1993) 321
\item{12.} V. Jain The 3D Effective Field Theory of the High Temperature
Abelian Higgs Model, MPI-Ph/92-72
\item{13.} V. Jain and A. Papadopolous, Phys. Lett. {303B} (1993) 315
\item{14.} I. Moss, D. Toms and A. Wright, Phys. Rev. {\bf D46} (1992) 1671
\item{15.} R.L. Stratonovich, Dokl. Akad. Nauk SSSR {\bf 115} (1957)\\
J. Hubbard, Phys. Rev. Lett. {\bf 3} (1959) 77

\end